\documentclass[12pt,twoside]{iopart}
\usepackage[latin9]{inputenc}
\usepackage{graphicx}
\usepackage{esint}

\makeatletter
\usepackage{iopams}
\usepackage{setstack}

\usepackage{iopams}
\eqnobysec

\makeatother

\begin{document}

\title{Loop exponent in DNA bubble dynamics}

\author{Vojt\v{e}ch Kaiser$^{1,2,3,*}$ and Tom\'{a}\v{s} Novotn\'{y}$^{1,\#}$}

\address{$^{1}$Department of Condensed Matter Physics, Faculty of Mathematics
and Physics, Charles University in Prague, Ke Karlovu 5, 121 16 Prague,
Czech Republic}

\address{$^{2}$Laboratoire de Physique, \'{E}cole Normale Sup\'{e}rieure de Lyon,
Universit\'{e} de Lyon, 46~All\'{e}e d'Italie, 69364 Lyon Cedex 07, France}

\address{$^{3}$Max-Planck-Institut f\"{u}r Physik komplexer Systeme, N\"{o}thnitzer
Stra{\ss}e 38, 01187 Dresden, Germany}

\ead{$ $$*$vojtech.kaiser@ens-lyon.fr; $\#$tno@karlov.mff.cuni.cz}
\begin{abstract}
Dynamics of DNA bubbles are of interest for both statistical physics
and biology. We present exact solutions to the Fokker-Planck equation
governing bubble dynamics in the presence of a long-range entropic
interaction. The complete meeting time and meeting position probability
distributions are derived from the solutions. Probability distribution
functions reflect the value of the loop exponent of the entropic interaction.
Our results extend previous results which concentrated mainly on the
tails of the probability distribution functions and open a way to
determining the strength of the entropic interaction experimentally
which has been a matter of recent discussions. Using numerical integration,
we also discuss the influence of the finite size of a DNA chain on
the bubble dynamics. Analogous results are obtained also for the case
of subdiffusive dynamics of a DNA bubble in a heteropolymer, revealing
highly universal asymptotics of meeting time and position probability
functions. 
\end{abstract}

\pacs{05.10.Gg, 82.37.j, 87.15.v, 87.14.gk}

\submitto{\JPA}
\maketitle

\section{Introduction}

DNA bubbles are local openings of the DNA double-helix caused by thermal
or torsional forces. The genetic information is stored inside of the
double-helix. Hence, the bubbles facilitate binding proteins to DNA
for transcription, replication, and repair \cite{Ambjornsson:BiophysJ07}.
Bubbles (also known as loops) influence DNA thermodynamics and mediate
long-range interactions along the DNA chain necessary for the existence
of the melting phase transition \cite{Peyrard:NatPhys06}.

DNA undergoes a melting (denaturation) transition during which the
double helix separates into two single strands. The nature of the
DNA melting transition was first described by Poland and Scheraga
\cite{PolandScheraga1966A,PolandScheraga1966B}. Their theory assumes
that a DNA molecule consists of loops and bound double-stranded segments.
The Poland-Scheraga model is a basic yet extendable model of DNA;
it constitutes a limiting case of more complex models which motivates
our study of its dynamics. 

In the Poland-Scheraga model, whose form of the Gibbs free energy
we use in this work, the strength of the long-range interactions and
thus the order of the phase transition is determined by the \emph{loop
entropy exponent} $c$ multiplying the logarithmic potential term.
The value of $c$ depends on the self-avoiding and mutually avoiding
properties of DNA bubbles. Diffusion in a logarithmic potential has
been the subject of recent studies \cite{Barkai:PRE12,Mukamel:JSTAT12}. 

Previous works studied the bubble dynamics in DNA denaturation using
a sequence-averaged continuous model \cite{HankeMetzler2003,Altan-Bonnet2003,FogedbyMetzler2007A,FogedbyMetzler2007B,Bar:JPCM09}.
Extending this model, we aim to show how the value of the loop exponent
could be determined from bubble-closing dynamics. We focus on the
physical aspects of the dynamics and do not explicitly discuss biological
processes involving DNA bubbles in this work, because recent studies
have shown them to be strongly sequence-dependent \cite{Ambjornsson:PRL06,Jeon:JCP06,Banik:JCP11}
and controlled by the interplay between torsional and thermal forces
\cite{BenhamBi2004,JeonAdamcik2010,Ma:Science13,King:PNAS13,Oberstrass:PRL13}.

\begin{figure}
\centering \includegraphics{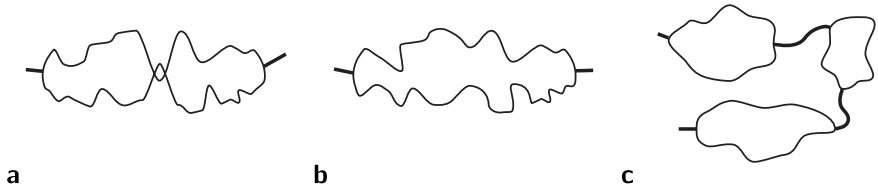}\label{fig:01} 

\protect\caption{\textbf{Loop entropy coefficient $c$ dependence on the excluded volume
interactions taken into account.} \textbf{a)} No exclusion effects,
$c=3/2$ comes purely from the constraint on the bubble to form a
closed loop \cite{PolandScheraga1966B}. \textbf{b)} Excluded volume
within a single bubble, $c=1.76$ given by the statistics of 3D self-avoiding
walks \cite{Fisher1966}. \textbf{c)} Excluded volume among all segments
of the DNA chain, renormalized value is $c=2.12$ \cite{KafriMukamel2000}.}
\end{figure}

Fisher \cite{Fisher1966} obtained $c=1.76$ considering only excluded-volume
interactions among bases in the same bubble, corresponding to a continuous
phase transition. This approximation is valid for short chains ($\sim10^{4}$
base pairs) \cite{HankeMetzler2003}. If inter-bubble and inter-chain
interactions are included as well, a calculation based on the polymer
network theory predicts $c=2.12$ in the thermodynamic limit, yielding
a first-order phase transition \cite{KafriMukamel2000} (see Fig.
\ref{fig:01} for an illustration of the excluded-volume interactions).

The value of the loop entropy exponent has not yet been conclusively
measured due to its strong correlation with the \emph{cooperativity
parameter} $\sigma$ \cite{BlosseyCarlon2003}. Therefore, we study
how $c$ influences DNA bubble closing dynamics, which are independent
of $\sigma$. Previous results \cite{FogedbyMetzler2007A,FogedbyMetzler2007B,Bar:JPCM09,Wu:PRE09}
focused mainly on the long-time asymptotics where we expect the experimental
signal-to-noise ratio to be low. The solution presented below is valid
for all times. Moreover, we also find the solution for the meeting
position, which especially around its peak proves as a useful complementary
quantity for experimental verification. 

We study two regimes of bubble dynamics -- \emph{diffusive} and \emph{subdiffusive}
-- shown to exist for a random DNA sequence \cite{HwaMarinari2003}.
The subdiffusive regime appears in the continuous limit of bubble
trapping in AT rich regions, whose melting is energetically less costly
than GC-region melting. For both of these regimes, we compute the
probability distribution functions (PDFs) of the meeting time, i.e.,
the time when the bubble ends meet and the bubble closes, and the
meeting position PDFs. The computations are carried out analytically
in the Laplace picture and numerically inverse transformed to the
time domain afterwards.

In the diffusive regime, we show how the loop exponent influences
bubble dynamics in a finite DNA chain. Our result extends previous
results for finite molecules that neglected the loop entropy\cite{Bicout:PRE04,NovotnyPedersen2007,PedersenHansen2009}.
Meeting time and meeting position PDFs are obtained using the finite-element
method.

The article is organised as follows. Section \ref{sec:Model-definition}
details our model. Section \ref{sec:Diffusive-dynamics} solves the
diffusive model, while section \ref{sec:Subdiffusive-dynamics} the
subdiffusive one. In section \ref{sec:Finite-chains}, we describe
bubble diffusion along finite DNA chains. In section \ref{sec:Experimental-verification},
we briefly comment on the experimental feasibility of verifying the
obtained results based on known and estimated thermodynamic parameters
of DNA. We close with concluding section \ref{sec:Conclusions}.

\section{Model definition\label{sec:Model-definition}}

The Poland-Scheraga model describes the DNA molecule as a one-dimensional
chain of alternating parts: loops and bound segments \cite{PolandScheraga1966B,KafriMukamel2002}.
Initiating a loop and denaturing a subsequent base pair are associated
with their Gibbs free energy $G_{\mathrm{init}}$ and $\Delta G_{\mathrm{bp}}$,
respectively. The value of $\Delta G_{\mathrm{bp}}$ is usually determined
by melting small DNA molecules \cite{SantaLuciaHicks2004} and corresponds
to a complete denaturation of the DNA chain. A closed loop can attain
fewer configurations than two fully denatured strands. The difference
in allowed configurations scales as a power of the number of open
base pairs in the loop with the loop entropy exponent $c$. The entropy
reduction is thus logarithmic in the bubble size: $\Delta S_{\mathrm{loop}}=-k_{\mathrm{B}}\, c\ln n$.

The dependence of entropy on the loop length mediates an effective
long-range interaction between the ends of the loop where the opening
and closing dynamics take place. Because the entropies of individual
loops contribute to the total Gibbs energy $G=H-TS$ of the DNA molecule,
the system interacts on a long range, which is a necessary condition
for a phase transition to occur in one dimension \cite{HouchesLRI}.
The long-range interactions grow more prominent with increasing $c$.
The transition occurs if $c>1$ and is of the first order for $c>2$.

\subsection{Gibbs free energy landscape}

Initiating a loop costs a substantial amount of energy ($G_{\mathrm{init}}\simeq10k_{\mathrm{B}}T_{\mathrm{phys}}$
at physiological temperatures $T_{\mathrm{phys}}\simeq310$ K). The
statistical weight of opening the bubble $\sigma(T)=\exp(-\beta G_{\mathrm{init}})$
is called the cooperativity parameter; its typical value is $\sigma(T_{\mathrm{phys}})\simeq10^{-4}$.
Bubbles are thus separate thermal excitations except close to the
melting temperature $T_{\mathrm{c}}$ \cite{HankeMetzler2003}. Treating
a single bubble approximates the DNA chain dynamics well. 

\begin{figure}[ht]
\centering \includegraphics{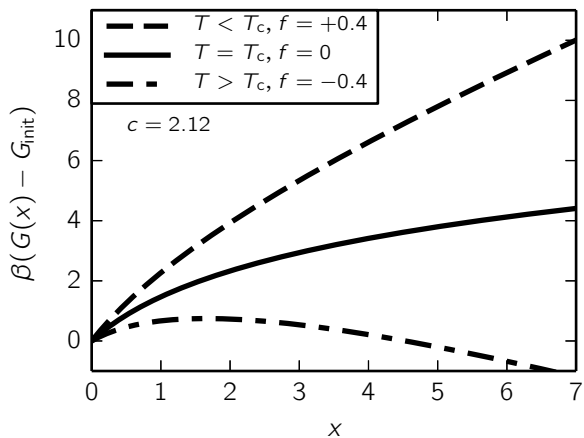} \protect\caption{\textbf{Character of the Gibbs free energy (in units of $k_{\mathrm{B}}T$)
as a function of the number of open base pairs for various temperatures}.
Under the melting temperature $T_{\mathrm{c}}$, both the linear term
$fx=\beta\Delta G_{bp}x/2$ and the logarithmic term $c/x$ contribute
to the closing of the bubble. At the phase transition, only the logarithmic
term remains. Above $T_{\mathrm{c}}$, the logarithmic term serves
as a nucleation barrier to the denaturation of DNA.}
\label{fig:02} 
\end{figure}

We approximate the number of open base pairs $n$ by a continuous
variable $x\geq0$. The Gibbs free energy for a single bubble with
$x$ open base pairs depicted in Figure \ref{fig:02} then reads:
\begin{equation}
G(x)=G_{\mathrm{init}}+x\Delta G_{\mathrm{bp}}+k_{\mathrm{B}}T\, c\ln x\,.\label{ContinuousGibbsFE}
\end{equation}

\subsection{Two types of DNA bubble dynamics}

Bubbles in a random (spatially uncorrelated) DNA sequence have two
regimes of propagation separated by a glass transition: diffusive
and subdiffusive\ \cite{HwaMarinari2003}. Below the glass transition
($T\leq T_{\mathrm{g}}\leq T_{\mathrm{c}}$), the dynamics of bubbles
are diffusive. Between $T_{\mathrm{g}}$ and $T_{\mathrm{c}}$, the
subdiffusive regime appears with an exponent of subdiffusion $0<\nu<1$
monotonously decreasing from unity at $T_{\mathrm{g}}$ to zero at
$T_{\mathrm{c}}$.

We consider only the closing dynamics of a single bubble in the following,
due to the low statistical weight $\sigma(T)\ll1$ for reopening of
the bubble. The dynamics being dominated by loop closing was observed
experimentally in \cite{Altan-Bonnet2003}.

\section{Diffusive dynamics\label{sec:Diffusive-dynamics}}

\subsection{Bubble-size Fokker-Planck equation}

In the continuous approximation, the Fokker-Planck equation (FPE)
describes the diffusive dynamics of a bubble. Given a potential $G$,
FPE has the form\ \cite[Sec.~5.4]{Risken1989}: 
\begin{equation}
\fl\partial_{\tau}P_{\mathrm{s}}(x,\tau|x_{0})=-\partial_{x}S_{\mathrm{s}}(x,\tau|x_{0})=K\partial_{x}\left[\partial_{x}P_{\mathrm{s}}(x,\tau|x_{0})+\beta P_{\mathrm{s}}(x,\tau|x_{0})\partial_{x}G(x)\right]\;,\label{FPEgeneral}
\end{equation}
where $P_{\mathrm{s}}(x,\tau|x_{0})$ is the bubble-size PDF%
\footnote{In the text we use $P_{\mathrm{s}}$, $P_{\mathrm{c}}$, and $P_{\mathrm{e}}$
to denote the bubble size, centre, and edge PDFs, respectively.%
} at time $\tau$, $S_{\mathrm{s}}(x,\tau|x_{0})$ the probability
current, $K$ the diffusion constant, and $\beta$ the inverse temperature.
Initially, $x_{0}$ base pairs are open and $P_{\mathrm{s}}(x,t=0|x_{0})=\delta(x-x_{0})$.
We introduce dimensionless time $t=\tau K/2$, where $K$ can be regarded
as a fitting parameter for comparison with the experiment. The definition
of dimensionless time $t$ reflects that the bubble shrinks or grows
at its edges, which propagate with diffusion constant $K/2$.

The potential \eref{ContinuousGibbsFE} introduces a drift term
opening or closing the bubble 
\begin{equation}
\partial_{x}G(x)=\Delta G_{bp}+k_{B}Tc/x\,.\label{DriftForce}
\end{equation}
Substituting \eref{DriftForce} into \eref{FPEgeneral} yields
\begin{equation}
\partial_{t}P_{\mathrm{s}}(x,t|x_{0})-2\partial_{xx}P_{\mathrm{s}}(x,t|x_{0})-4\partial_{x}\left[\left(f+\frac{\gamma}{x}\right)P_{\mathrm{s}}(x,t|x_{0})\right]=0\,,\label{FPEreduced}
\end{equation}
where $f\equiv\beta\Delta G_{bp}/2$ and $\gamma\equiv c/2$ are introduced.
The absorbing boundary condition {$\lim_{x\rightarrow0+}P_{\mathrm{s}}(x,t|x_{0})=0$}
represents closing dynamics. We focus on the regime $T<T_{\mathrm{c}}$,
where $f>0$.

\subsection{Mapping to the Coulomb problem}

FPE \eref{FPEreduced} maps to the imaginary-time Coulomb problem
\cite{FogedbyMetzler2007A,FogedbyMetzler2007B}. This mapping is an
example of transforming the Fokker-Planck operator to a Hermitian
one\ \cite{Risken1989}. The transformed PDF $w(x,t|x_{0})$ is related
to the original bubble size PDF by 
\begin{equation}
P_{\mathrm{s}}(x,t|x_{0})=e^{-f(x-x_{0})}\left(\frac{x}{x_{0}}\right)^{-\gamma}w(x,t|x_{0})\,.\label{PDFtransform}
\end{equation}
The spatial part of FPE \eref{FPEreduced} is then Hermitian 
\begin{eqnarray}
\left[-\partial_{t}+2\partial_{xx}-2f^{2}-\frac{4f\gamma}{x}-\frac{2\gamma(\gamma+1)}{x^{2}}\right]w(x,t|x_{0})=0\label{HermitianForm}\\
w(x,t\rightarrow0^{+}|x_{0})=\delta(x-x_{0}),\qquad w(x\rightarrow0^{+},t|x_{0})=0\,.\nonumber 
\end{eqnarray}
Fogedby and Metzler used this transform to obtain the spectrum, the
long-time asymptotics of bubble dynamics, as well as the complete
solution at the critical point where $f=0$\ \cite{FogedbyMetzler2007A,FogedbyMetzler2007B}.

\subsection{Green function}

While the spectral approach to \eref{HermitianForm} provides an
insight to the long-time behaviour of bubble dynamics, summing the
eigenfunctions to obtain the short-time dynamics proves to be difficult.
To obtain an exact Laplace picture form of the solution (denoted by
the bar over the quantities), we take a Laplace transform of \eref{HermitianForm}
instead: 
\begin{eqnarray}
\fl\left[\partial_{xx}-\left(f^{2}+\frac{s}{2}\right)-\frac{2f\gamma}{x}-\frac{\gamma(\gamma+1)}{x^{2}}\right]\bar{w}(x,s|x_{0})=-\frac{1}{2}\delta(x-x_{0})\;,\label{LaplaceForm1}\\
\bar{w}(x\to0^{+},s|x_{0})=0\,.\nonumber 
\end{eqnarray}
The task is to find the Green function of a differential operator
with $s$ as a parameter. We find homogeneous solutions of \eref{LaplaceForm1}
and construct the Green function as their linear combination.

The homogeneous solutions are the Whittaker functions $M_{(-f\gamma/\alpha(s);\gamma+1/2)}(2x\alpha(s))$
and $W_{(-f\gamma/\alpha(s);\gamma+1/2)}(2x\alpha(s))$, where we
defined $\alpha^{2}(s)=f^{2}+s/2$. $M_{(-f\gamma/\alpha(s);\gamma+1/2)}(2x\alpha(s))$
is regular for $x\rightarrow0$ and diverges at infinity; $W_{(-f\gamma/\alpha(s);\gamma+1/2)}(2x\alpha(s))$
is a regular solution at infinity and singular at the origin \cite{NISTfunctions2010}.

The Green function is continuous and has a first-derivative discontinuity
$\lim_{x\rightarrow x_{0}+}\partial_{x}\bar{w}(x,s|x_{0})-\lim_{x\rightarrow x_{0}-}\partial_{x}\bar{w}(x,s|x_{0})=-1/2$.
The linear combination of Whittaker functions that satisfies both
\eref{LaplaceForm1} and the above conditions is 
\begin{eqnarray}
\fl\bar{w}(x,s|x_{0})= & \frac{\Gamma(1+\gamma+\frac{f\gamma}{\alpha(s)})}{4\alpha(s)\Gamma(2+2\gamma)}\bigg[\theta(x-x_{0})M_{(-\frac{f\gamma}{\alpha(s)};\gamma+\frac{1}{2})}(2x_{0}\alpha(s))W_{(-\frac{f\gamma}{\alpha(s)};\gamma+\frac{1}{2})}(2x\alpha(s))+\nonumber \\
 & +\theta(x_{0}-x)W_{(-\frac{f\gamma}{\alpha(s)};\gamma+\frac{1}{2})}(2x_{0}\alpha(s))M_{(-\frac{f\gamma}{\alpha(s)};\gamma+\frac{1}{2})}(2x\alpha(s))\bigg]\,,
\end{eqnarray}
where the prefactor ensures the correct discontinuity of the first
derivative at $x_{0}$ and can be related to the constant Wronskian
of the Whittaker functions \cite[Sec.~13.14.26]{NISTfunctions2010}

\begin{equation}
\mathcal{W}\equiv M_{\kappa;\mu}(x)\frac{d}{dx}W_{\kappa;\mu}(x)-W_{\kappa;\mu}(x)\frac{d}{dx}M_{\kappa;\mu}(x)=-\frac{\Gamma(1+2\mu)}{\Gamma(\frac{1}{2}+\mu-\kappa)}.
\end{equation}

The Green function of \eref{FPEreduced} in the Laplace picture
is obtained by substitution of $\bar{w}(x,s|x_{0})$ into \eref{PDFtransform}
\begin{eqnarray}
\fl\bar{P}_{\mathrm{s}}(x,s|x_{0})= & e^{-f(x-x_{0})}\left(\frac{x}{x_{0}}\right)^{-\gamma}\frac{\Gamma(1+\gamma+\frac{f\gamma}{\alpha(s)})}{4\alpha(s)\Gamma(2+2\gamma)}\times\nonumber \\
 & \times\bigg[\theta(x-x_{0})W_{(-\frac{f\gamma}{\alpha(s)};\gamma+\frac{1}{2})}(2x\alpha(s))M_{(-\frac{f\gamma}{\alpha(s)};\gamma+\frac{1}{2})}(2x_{0}\alpha(s))+\nonumber \\
 & +\theta(x_{0}-x)M_{(-\frac{f\gamma}{\alpha(s)};\gamma+\frac{1}{2})}(2x\alpha(s))W_{(-\frac{f\gamma}{\alpha(s)};\gamma+\frac{1}{2})}(2x_{0}\alpha(s))\bigg]\,.\label{FullGreenFunction}
\end{eqnarray}
To our best knowledge, an analytical form of the inverse Laplace transform
of the previous equation does not exist and has to be evaluated numerically
instead. Current experimental techniques cannot directly observe time
dependence of the bubble-size PDF. Therefore, it is necessary to evaluate
derived PDFs, namely for the meeting time and meeting position. In
the following, we give the Laplace picture expressions for these PDFs.

\subsection{Meeting time PDF}

The probability of the bubble surviving up to time $t$ is the total
integral of the $P_{\mathrm{s}}(x,t|x_{0})$ over all possible $x$.
This can only be reduced by the probability flux out of the system
at its boundaries \cite{Redner2001,PedersenHansen2009,Novotny:PRE00},
which for our case happens if the two edges of a bubble meet. Thus,
we define the meeting time PDF $\bar{\pi}_{\mathrm{mt}}(t|x_{0})$
as the probability current through the absorbing boundary at the origin
\begin{equation}
\pi_{\mathrm{mt}}(t|x_{0})\equiv-S_{\mathrm{s}}(0,t|x_{0})=\lim_{x\rightarrow0}\left[2\partial_{x}+4\left(\frac{\gamma}{x}+f\right)\right]P_{\mathrm{s}}(x,t|x_{0})\,.
\end{equation}
After substituting from \eref{FullGreenFunction} into the previous
equation, the Laplace transform of the meeting time PDF reads 
\begin{eqnarray}
\fl\bar{\pi}_{\mathrm{mt}}(s|x_{0})= & \frac{e^{fx_{0}}}{\Gamma(1+2\gamma)}\left(x_{0}\sqrt{4f^{2}+2s}\right)^{\gamma}\Gamma\left(1+\gamma+f\gamma\sqrt{\frac{2}{2f^{2}+s}}\right)\times\nonumber \\
 & \times W_{\left(-f\gamma\sqrt{\frac{2}{2f^{2}+s}};\gamma+\frac{1}{2}\right)}\left(x_{0}\sqrt{4f^{2}+2s}\right)\,.\label{DiffusiveMeetingTime}
\end{eqnarray}

We invert the Laplace transform numerically. Generally, the numerical
inverse Laplace transform is ill-conditioned because the associated
operator is unbounded \cite{AbateWhitt2006}. Available numerical
methods restrict the operator to smaller spaces of functions. A class
of these methods invert functions with known pole structure, while
other methods operate on real axis both in the direct and Laplace
picture (see Refs.~\cite{AbateWhitt2006,HupperPollak1999} for an
overview).

We used the Stehfest algorithm \cite{Stehfest1970} in this work which
computes the Laplace inverse as a weighted sum of $\bar{\pi}_{\mathrm{mt}}(s|x_{0})$
for real $s$. The algorithm requires the resulting functions to be
continuous with bounded derivatives. The bounds of derivatives determine
the stability of the algorithm. In our case, the selected parameters
result in curves with sufficiently small derivatives. To further avoid
instabilities, we used high-precision arithmetics (up to 100 digits). 

\begin{figure}[ht]
\centering \includegraphics{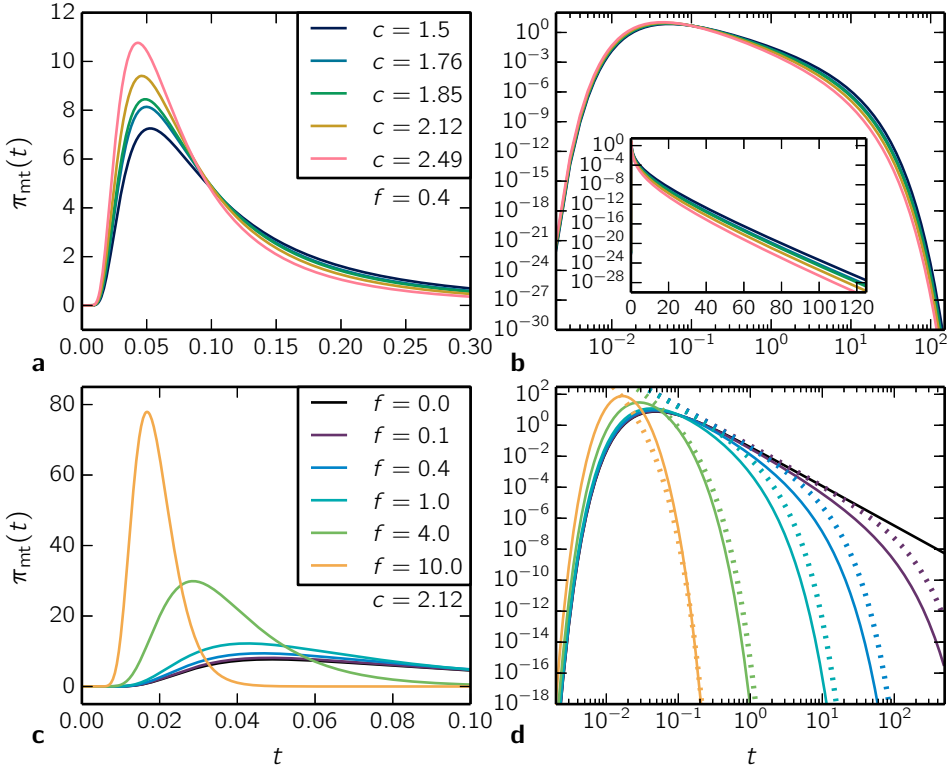} \protect\caption{\textbf{Meeting time PDFs. a,b)} Dependence of the meeting time PDFs
$\pi_{\mathrm{mt}}(t)$ on $c$ for $f=0.4$ and $x_{0}=1$. \textbf{c,d)}
Dependence of the meeting time PDFs $\pi_{\mathrm{mt}}(t)$ on $f$
for $c=2.12$ and $x_{0}=1$. The asymptotic behaviour exhibits a
transition from power law controlled by $c$ to an exponential decay
given by $f^{2}$, except for $T=T_{\mathrm{c}}$ (corresponding to
$f=0$), where the power law holds for all times. Dotted lines are
the asymptotic functions (\ref{MeetingTimeAsymptotics}) from Ref.~\cite{FogedbyMetzler2007A}.}
\label{fig:03} 
\end{figure}

The results are shown in Figure \ref{fig:03} together with the asymptotics
\begin{equation}
\pi_{\mathrm{mt}}\propto t^{-(c+3)/2}\exp(-2f^{2}t)\label{MeetingTimeAsymptotics}
\end{equation}
from Ref.~\cite{FogedbyMetzler2007A}. Both parameters $c$ and $f$
influence strongly not only the tails of the PDFs but also their peak
values.%
\footnote{The values in the figures correspond to an ideal loop ($c=1.5$),
self-avoiding loop ($c=1.76$) \cite{Fisher1966}, renormalised value
in the presence of an external stretching force ($c=1.85$) \cite{Hanke:PRL08,Marenduzzo:PRE10},
self-avoiding and mutually avoiding loops ($c=2.12$) \cite{KafriMukamel2000},
and the maximal allowed value for RNA with pseudoknots and hairpins
($c=2.49$) \cite{Netz:PRL08}.%
}

\subsection{Meeting position PDF}

Apart from changing its size, a bubble diffuses freely along the DNA
chain, described by the bubble centre position $y$ and its PDF $P_{\mathrm{c}}(y,t|y_{0})$.
The energy landscape is flat because pairing energies are sequence
averaged. The Fokker-Planck equation for the centre position reads
\begin{equation}
\partial_{t}P_{\mathrm{c}}(y,t|y_{0})=\partial_{yy}P_{\mathrm{c}}(y,t|y_{0})\;,\label{FPEcentre}
\end{equation}
where $y_{0}$ is the initial centre position. Without loss of generality,
we set $y_{0}=0$. The bubble position PDF is 
\begin{equation}
P_{\mathrm{c}}(y,t|0)=\frac{1}{\sqrt{2\pi t}}\exp{\left(-\frac{y^{2}}{2t}\right)}\,.\label{Gaussian}
\end{equation}

Integrating the product of the meeting time PDF and the bubble position
PDF over all times yields the meeting position PDF 
\begin{equation}
\pi_{\mathrm{mp}}(y|x_{0},y_{0})=\int_{0}^{\infty}\mathrm{d}t\,\pi_{\mathrm{mt}}(t|x_{0})P_{\mathrm{c}}(y,t|y_{0})\,.\label{MeetingPositionDef}
\end{equation}

The time when the bubble edges meet is projected onto the meeting
position which leads to the equivalence of the Fourier transform of
the meeting position PDF and the Laplace transform of the meeting
time PDF. We refer to this equivalence as the \emph{projection principle}
because closing dynamics of the bubble in the time variable $t$ are
completely projected onto the closing dynamics in the spatial variable
$y$.

The Fourier transform of the definition of the meeting position distribution
\eref{MeetingPositionDef} (denoted by tilde) reads 
\begin{equation}
\mathcal{F}\left[\pi_{\mathrm{mp}}\right](k|x_{0})\equiv\tilde{\pi}_{\mathrm{mp}}(k|x_{0})=\int_{0}^{\infty}\mathrm{d}t\,\pi_{abs}(t|x_{0})\tilde{P}_{c}(k,t)\,,\label{MeetingPositionDefFourier}
\end{equation}
where we use the unitary Fourier transform $\mathcal{F}\left[h(y)\right](k)=(2\pi)^{-1/2}\int_{-\infty}^{\infty}h(y)\exp(-iky)\mathrm{d}y$.
The Fourier transform of the probability distribution function of
the position of the bubble-centre PDF is given by 
\begin{equation}
\tilde{P}_{c}(k,t)=\frac{1}{\sqrt{2\pi}}e^{-\frac{k^{2}}{2}t}\,.
\end{equation}
Inserting this expression into \eref{MeetingPositionDefFourier}
yields 
\begin{equation}
\tilde{\pi}_{\mathrm{mp}}(k|x_{0})=\frac{1}{\sqrt{2\pi}}\int_{0}^{\infty}\mathrm{d}t\,\pi_{\mathrm{mt}}(t|x_{0})e^{-\frac{k^{2}}{2}t}\,.
\end{equation}

Finally, we compare the previous relation with the Laplace transform
of the meeting time probability density 
\begin{equation}
\mathcal{L}\left[\pi_{\mathrm{mt}}(t|x_{0})\right](s|x_{0})\equiv\bar{\pi}_{\mathrm{mt}}(s|x_{0})=\int_{0}^{\infty}\mathrm{d}t\,\pi_{\mathrm{mt}}(t|x_{0})e^{-st}\,,
\end{equation}
which leads to the relation 
\begin{equation}
\tilde{\pi}_{\mathrm{mp}}(k|x_{0})=\frac{1}{\sqrt{2\pi}}\bar{\pi}_{\mathrm{mt}}(s|x_{0})\Big|_{s\rightarrow\frac{k^{2}}{2}}\,.
\end{equation}
This is the explicit form of the projection principle described above.

\begin{figure}[ht]
\centering \includegraphics{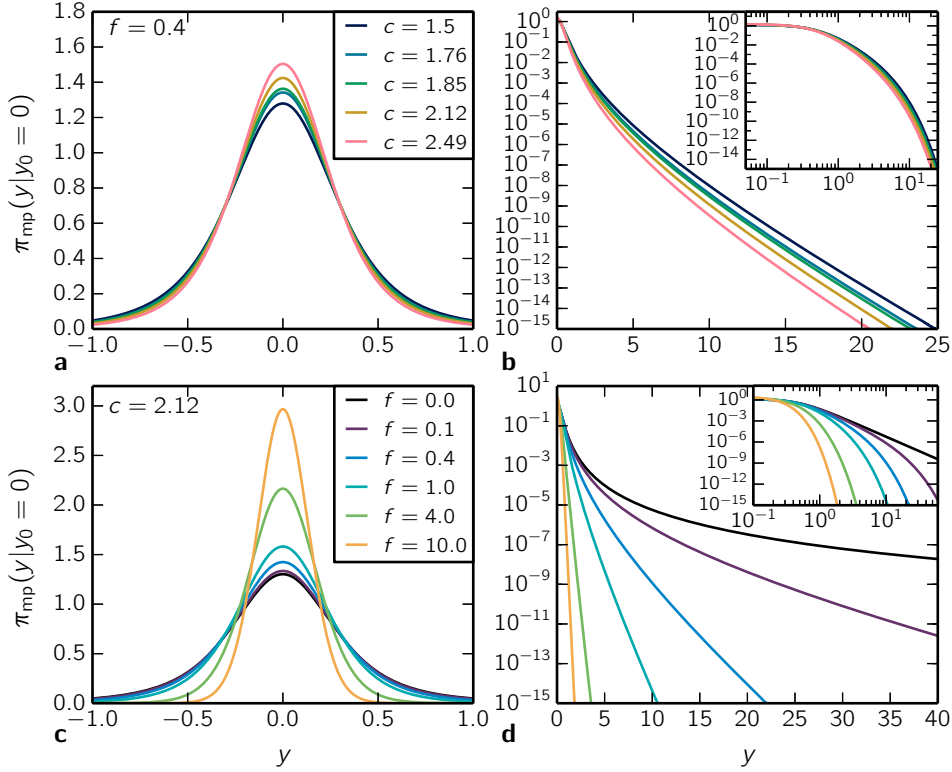} \protect\caption{\textbf{Meeting position PDFs.} \textbf{a,b)} Dependence of the meeting
position PDFs on $c$ for $f=0.4$ and $x_{0}=1$, $y_{0}=0$. \textbf{c,d)}
Dependence of the meeting position PDFs on $f$ for $c=2.12$ and
$x_{0}=1$, $y_{0}=0$. The asymptotics show a transition from the
power law behaviour at the melting temperature ($f=0$) to an exponential
decay for finite $f$. This is similar to the case of the meeting
time PDF, although the onset of the exponential behaviour is more
pronounced for the meeting position.}
\label{fig:04} 
\end{figure}

The Fourier picture expression for the meeting position PDF is 
\begin{eqnarray}
\fl\tilde{\pi}_{\mathrm{mp}}(k|x_{0})= & \frac{1}{\sqrt{2\pi}}\frac{e^{fx_{0}}}{\Gamma(1+2\gamma)}\left(x_{0}\sqrt{4f^{2}+k^{2}}\right)^{\gamma}\Gamma\left(1+\gamma+\frac{2f\gamma}{\sqrt{4f^{2}+k^{2}}}\right)\times\nonumber \\
 & \times W_{\left(-\frac{2f\gamma}{\sqrt{4f^{2}+k^{2}}};\gamma+\frac{1}{2}\right)}\left(x_{0}\sqrt{4f^{2}+k^{2}}\right)\,.
\end{eqnarray}
Its inverse Fourier transform can be obtained numerically using Fast
Fourier transform (FFT) based algorithms \cite[Sec.~13.8]{PressTeukolsky2007}.
For $f=0$, the result can be expressed analytically in the form of
a modified Lorenz distribution 
\begin{equation}
\pi_{\mathrm{mp}}(y|x_{0},y_{0})=\frac{2x_{0}^{1+c}}{\sqrt{\pi}}\frac{\Gamma\left(\frac{c}{2}+1\right)}{\Gamma\left(\frac{c}{2}+\frac{1}{2}\right)}\left(x_{0}^{2}+4(y-y_{0})^{2}\right)^{-\left(1+\frac{c}{2}\right)}\,.
\end{equation}
For non-zero $f$, we again observe an exponential suppression of
the PDF tails (Figure \ref{fig:04}).

\section{Subdiffusive dynamics\label{sec:Subdiffusive-dynamics}}

\subsection{Fractional Fokker-Planck equation}

Subdiffusion can be described similarly to diffusion, by introducing
fractional time derivatives. The fractional Fokker-Planck equation
(FFPE) is the subdiffusive analogue of the diffusive FPE \eref{FPEgeneral}.
Introduction to and mathematical background of subdiffusion and FFPE
is given in reviews \cite{MetzlerKlafter2000,MetzlerKlafter2004}.
The bubble-size FFPE reads 
\begin{eqnarray}
\partial_{\tau}P_{\mathrm{s}}^{(\nu)}(x,\tau|x_{0}) & =-\partial_{x}S_{\mathrm{s}}^{(\nu)}(x,\tau|x_{0})\label{FFPEbubbleSize}\\
 & =\ _{0}\mathcal{D}_{\tau}^{1-\nu}K_{\nu}\partial_{x}\left[\partial_{x}P_{\mathrm{s}}^{(\nu)}(x,\tau|x_{0})+\beta P_{\mathrm{s}}^{(\nu)}(x,\tau|x_{0})\partial_{x}G(x)\right],\nonumber 
\end{eqnarray}
where $\ _{0}\mathcal{D}_{\tau}^{1-\nu}$ is the Riemann-Liouville
differential operator \cite{MetzlerKlafter2000} 
\begin{equation}
_{0}\mathcal{D}_{\tau}^{1-\nu}f(\tau)\equiv\frac{\partial_{\tau}}{\Gamma(\nu)}\int_{0}^{\tau}\mathrm{d}\tau'\frac{f(\tau')}{(\tau-\tau')^{1-\nu}}\,.
\end{equation}
Solutions of FFPE are \emph{subordinate} to solutions of the corresponding
FPE \cite{MetzlerKlafter2004} 
\begin{equation}
\bar{P}_{\mathrm{s}}^{(\nu)}(x,s|x_{0})=s^{\nu-1}\bar{P}_{\mathrm{s}}(x,s^{\nu}|x_{0})\,.\label{RelationSubDiff}
\end{equation}
A similar relation exists for the current. Comparing \eref{FPEgeneral}
and \eref{FFPEbubbleSize} gives 
\begin{equation}
\bar{S}_{\mathrm{s}}^{(\nu)}(x,s|x_{0})=\bar{S}_{\mathrm{s}}(x,s^{\nu}|x_{0})\,.\label{RelationSubDiffCurrent}
\end{equation}

\subsection{Subdiffusive Green function}

Since the transform \eref{PDFtransform} only acts on the spatial
part of the bubble-size FPE, analogy of \eref{HermitianForm} can
be found for subdiffusion, with dimensionless time $t=\tau[K_{\nu}/2]^{1/\nu}$
\begin{eqnarray}
\fl-\partial_{t}w^{(\nu)}(x,t|x_{0})+\ _{0}\mathcal{D}_{t}^{1-\nu}\left[2\partial_{xx}-2f^{2}-\frac{4f\gamma}{x}-\frac{2\gamma(\gamma+1)}{x^{2}}\right]w^{(\nu)}(x,t|x_{0})=0\label{HermitianFormSD}\\
w^{(\nu)}(x,t\rightarrow0^{+}|x_{0})=\delta(x-x_{0}),\qquad w^{(\nu)}(x\rightarrow0^{+},t|x_{0})=0\,.\nonumber 
\end{eqnarray}
Using \eref{RelationSubDiff}, we find the corresponding subdiffusive
Green function 
\begin{eqnarray}
\fl\bar{P}_{\mathrm{s}}^{(\nu)}(x,s|x_{0})= & s^{\nu-1}e^{-f(x-x_{0})}\left(\frac{x}{x_{0}}\right)^{-\gamma}\frac{\Gamma(1+\gamma+\frac{f\gamma}{\alpha_{\nu}(s)})}{4\alpha_{\nu}(s)\Gamma(2+2\gamma)}\times\nonumber \\
 & \times\bigg[\theta(x-x_{0})W_{(-\frac{f\gamma}{\alpha_{\nu}(s)};\gamma+\frac{1}{2})}(2\alpha_{\nu}(s)x)M_{(-\frac{f\gamma}{\alpha_{\nu}(s)};\gamma+\frac{1}{2})}(2\alpha_{\nu}(s)x_{0})+\nonumber \\
 & +\theta(x_{0}-x)M_{(-\frac{f\gamma}{\alpha_{\nu}(s)};\gamma+\frac{1}{2})}(2\alpha_{\nu}(s)x)W_{(-\frac{f\gamma}{\alpha_{\nu}(s)};\gamma+\frac{1}{2})}(2\alpha_{\nu}(s)x_{0})\bigg]\,,\\
 & \mathrm{where}\qquad\alpha_{\nu}(s)=\sqrt{f^{2}+s^{\nu}/2}\,.\nonumber 
\end{eqnarray}

\subsection{Meeting time PDF\label{sub:Meeting-time-PDF}}

The subdiffusive meeting time PDF follows from substituting \eref{DiffusiveMeetingTime}
into \eref{RelationSubDiffCurrent} and has again the meaning of
the probability flow into the boundary due to the bubble closing events
\begin{eqnarray}
\fl\bar{\pi}_{\mathrm{mt}}^{(\nu)}(s|x_{0})= & \frac{e^{fx_{0}}}{\Gamma(1+2\gamma)}\left(x_{0}\sqrt{4f^{2}+2s^{\nu}}\right)^{\gamma}\Gamma\left(1+\gamma+f\gamma\sqrt{\frac{2}{2f^{2}+s^{\nu}}}\right)\times\nonumber \\
 & \times W_{\left(-f\gamma\sqrt{\frac{2}{2f^{2}+s^{\nu}}};\gamma+\frac{1}{2}\right)}\left(x_{0}\sqrt{4f^{2}+2s^{\nu}}\right)\,.
\end{eqnarray}

\begin{figure}[ht]
\centering \includegraphics{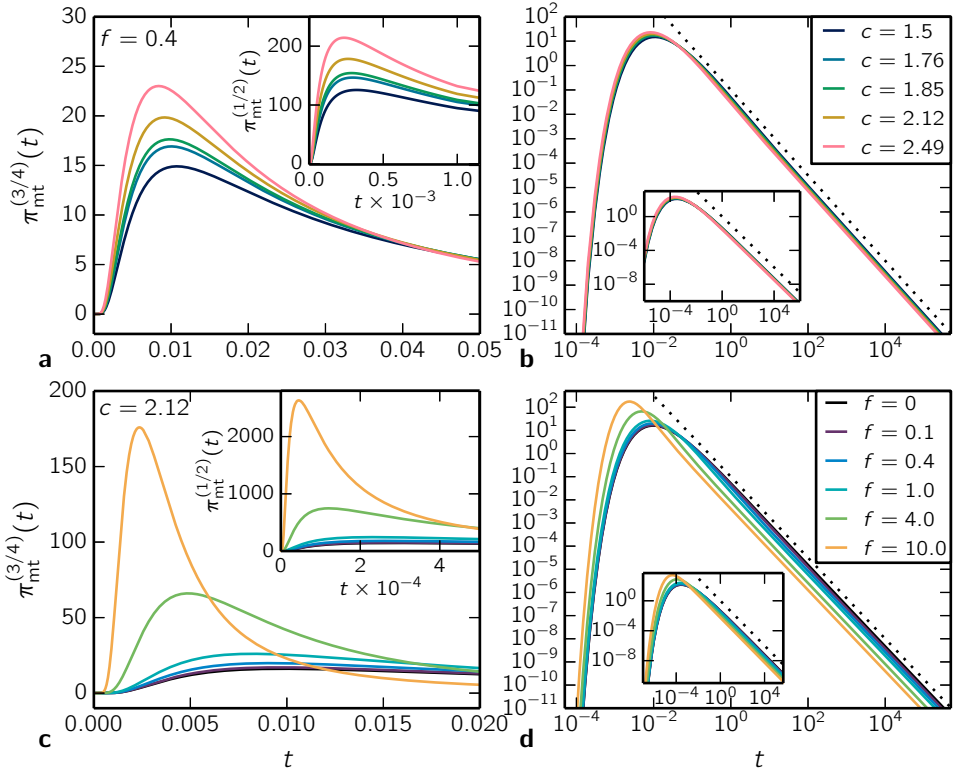} \protect\caption{\textbf{Meeting time PDFs for subdiffusion.} \textbf{a,b)} Dependence
of the meeting time PDFs on $c$ for $f=0.4$ and $x_{0}=1$. \textbf{c,d)}
Dependence of the meeting time PDFs on $f$ for $c=2.12$ and $x_{0}=1$.
The main figures show results for $\nu=3/4$ while the insets those
for $\nu=1/2$. The dotted lines show the universal power-law behaviour
with the exponent $-(1+\nu)$.}
\label{fig:05} 
\end{figure}

The inverse Laplace transform is obtained numerically as in the diffusive
case (Section \ref{sec:Diffusive-dynamics}) and plotted in Figure
\ref{fig:05}. We observe a universal behaviour for large times and
all values of $f$ and $c$ studied, where the meeting time PDFs decay
as a power law with exponent $-(1+\nu)$ (see Figure \ref{fig:05}).

We propose the following explanation of this effect. We assume the
existence of a finite mean meeting time 
\begin{equation}
T_{1}=-\lim_{s\rightarrow0^{+}}\frac{\bar{\pi}_{\mathrm{mt}}(s|x_{0})}{s}\,.\label{MeanMeetingTimeDef}
\end{equation}
of the diffusive meeting time PDF. Due to $\pi_{\mathrm{mt}}$ being
bounded and having asymptotics \eref{MeetingTimeAsymptotics}, the
finiteness of $T_{1}$ is guaranteed for all $c>0$ below the melting
temperature ($f>0$) and for $c>1$ at the transition ($f=0$). The
meeting time PDF can be expanded around $s=0$ as $\bar{\pi}_{\mathrm{mt}}(s|x_{0})=1-T_{1}s+\mathcal{O}(s^{2})$
in the Laplace picture. Subordination gives $\bar{\pi}_{\mathrm{mt}}^{(\nu)}(s|x_{0})=1-T_{1}s^{\nu}+\mathcal{O}(s^{2\nu})$
for small $s$ in the subdiffusive case. Tauberian theorems \cite[Ch.~XIII, Eq.~(5.22)]{Feller1971}
lead to $\pi_{\mathrm{mt}}^{(\nu)}(t\rightarrow\infty|x_{0})=t^{-(1+\nu)}+\mathcal{O}(t^{-(1+2\nu)})$
in the time domain for $\nu<1$. Alternatively, the scaling can be
derived by considering the subdiffusive dynamics to be a limit of
a continuous time random walk \cite{Lua:PRE05,Condamin:PRL07,Palyulin:JPA14}. 

In contrast to the universal behaviour of the tails, peaks of meeting
time PDFs show a very strong dependence on $f$ and $c$, which is
even more pronounced than in the diffusive case. The peak values of
the subdiffusive meeting time PDF surpass those in the diffusive case.
Returning to the physical origin of the subdiffusive behaviour allows
us to explain this behaviour. 

The subdiffusive regime appears as the heterogeneity of the DNA sequence
leads to the trapping of bubbles in AT-rich (soft) regions surrounded
by GC-rich regions \cite{HwaMarinari2003}. The hard-to-open GC-rich
regions limit the growth of the bubble, which can be likened to the
presence of a reflective boundary condition. The presence of a boundary
decreases the mean meeting time, as we observe below in the case of
finite DNA chains. However, if the bubble overcomes such a GC-rich
barrier, it spreads out to the next AT-rich region and its survival
time increases. This corresponds to the power law behaviour in tails
which results in a small fraction of long-lived bubbles which is larger
than in the diffusive case.

\begin{figure}[ht]
\centering \includegraphics{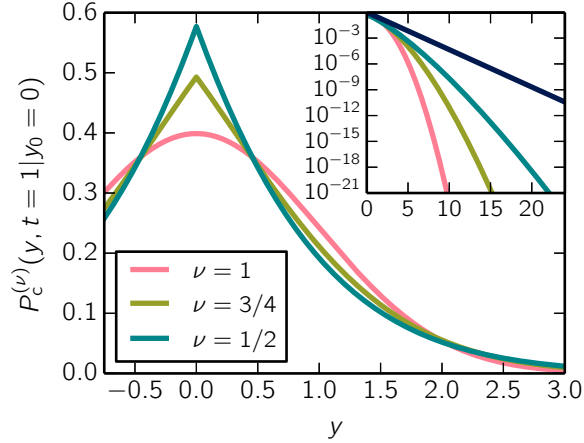} \protect\caption{\textbf{Bubble centre PDFs for free subdiffusion.} $P_{\mathrm{c}}^{(\nu)}(y,t=1|y_{0}=0)$
given by Eq.~(\ref{SubdiffBubblePositionPDF}) for selected values
of $\nu$ ($\nu=1$ corresponding to the standard diffusion (\ref{Gaussian})
is shown for comparison). The cusp at $y=0$ for $\nu<1$ is a characteristic
of the subdiffusive behaviour. Inset: large-distance asymptotics are
given by exponentials of powers (bounded between $1$ and $2$, compare
with Eqs.~(\ref{eq:Asympt1/2}) and (\ref{eq:Asympt3/4})) of $y$.
Dark blue topmost straight line shows a simple exponential for comparison.}
\label{fig:06} 
\end{figure}

\subsection{Meeting position PDF\label{sub:Meeting-position-PDF}}

The definition of the meeting position PDF \eref{MeetingPositionDef}
holds for subdiffusive dynamics as well. However, subdiffusive dynamics
do not lead to a simple Gaussian form of the bubble position PDF and,
therefore, the projection principle cannot be used. Subdiffusive bubble
position PDF solves the appropriately modified Eq.~(\ref{FPEcentre})
\begin{equation}
\partial_{t}P_{\mathrm{c}}^{(\nu)}(y,t|y_{0})={}_{0}\mathcal{D}_{t}^{1-\nu}\partial_{yy}P_{\mathrm{c}}^{(\nu)}(y,t|y_{0}),\label{SubdiffFPEcentre}
\end{equation}
 with the following scaling property of the solution $P_{\mathrm{c}}^{(\nu)}(y,t|y_{0})=P_{\mathrm{c}}^{(\nu)}((y-y_{0})/t^{\nu/2},1|0)/t^{\nu/2}$
stemming directly from the joint Laplace and Fourier transform of
Eq.~(\ref{SubdiffFPEcentre}). Explicitly, the solution is described
in terms of the Fox H-functions \cite{MetzlerKlafter2000} 
\begin{equation}
P_{\mathrm{c}}^{(\nu)}(y,t|y_{0})=\frac{1}{\sqrt{2\pi t^{\nu}}}H_{1,2}^{2,0}\left[\frac{(y-y_{0})^{2}}{2t^{\nu}}\bigg|\begin{array}{l}
(1-\frac{\nu}{2},\nu)\\
(0,1),(\frac{1}{2},1)
\end{array}\right]\,,\label{SubdiffBubblePositionPDF}
\end{equation}
which reduce to the Meijer G-functions (more commonly implemented
in numerical libraries than the Fox H-functions), if $\nu=q/p$ is
rational. However, the number of Meijer G-function parameters increases
with growing $q$ and $p$, increasing the computational effort for
evaluating the functions and decreasing their precision. Useful special
cases are $\nu=1/2$ and $\nu=3/4$, for which the bubble centre PDFs
take on the forms 
\begin{eqnarray}
P_{\mathrm{c}}^{(1/2)}(y,t|y_{0}) & =\frac{1}{2\pi^{\frac{3}{2}}t^{\frac{1}{4}}}G_{0,3}^{3,0}\left[\frac{(y-y_{0})^{4}}{64t}\bigg|\overline{(0,\frac{1}{4},\frac{1}{2})}\right]\,,\nonumber \\
P_{\mathrm{c}}^{(3/4)}(y,t|y_{0}) & =\frac{1}{2\sqrt{2}3^{\frac{1}{8}}\pi^{\frac{5}{2}}t^{\frac{3}{8}}}G_{2,7}^{7,0}\left[\frac{(y-y_{0})^{8}}{32^{4}t^{3}}\bigg|\begin{array}{l}
(\frac{5}{24},\frac{13}{24})\\
(0,\frac{1}{8},\frac{1}{4},\frac{3}{8},\frac{1}{2},\frac{5}{8},\frac{3}{4})
\end{array}\right]\,,
\end{eqnarray}
with asymptotics reading 
\begin{eqnarray}
P_{\mathrm{c}}^{(1/2)}(u\rightarrow\infty,1|0) & \propto\exp\left(-\frac{3}{4}u^{\frac{4}{3}}\right)\left(u^{-\frac{1}{3}}+\mathcal{O}\left(u^{-\frac{4}{3}}\right)\right)\,,\label{eq:Asympt1/2}\\
P_{\mathrm{c}}^{(3/4)}(u\rightarrow\infty,1|0) & \propto\exp\left(-\frac{3^{\frac{3}{5}}5}{16}u^{\frac{8}{5}}\right)\left(u^{-\frac{1}{5}}+\mathcal{O}\left(u^{-\frac{6}{5}}\right)\right)\,.\label{eq:Asympt3/4}
\end{eqnarray}
These functions, along with the Gaussian (\ref{Gaussian}) for comparison,
are plotted in Figure~\ref{fig:06} which shows their characteristic
non-analytic behaviour at zero (cusp, i.e. the discontinuity in first
derivative). The inset shows their localisation around zero (stronger
than exponential but weaker than Gaussian). 

\begin{figure}[ht]
\centering \includegraphics{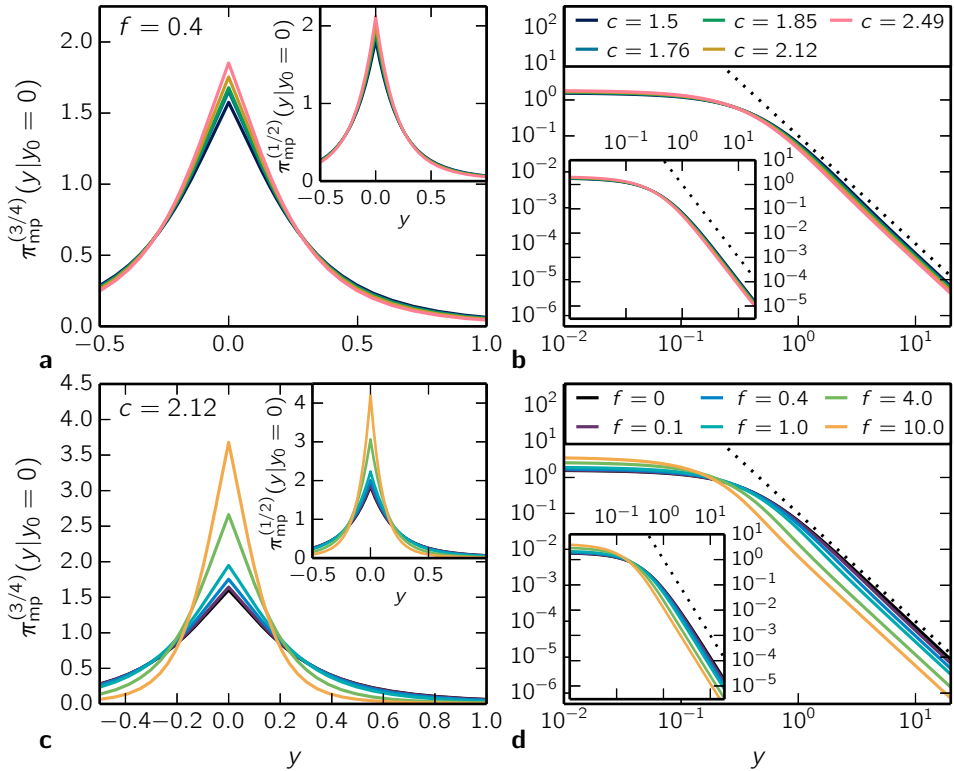} \protect\caption{\textbf{Meeting position PDFs for subdiffusion (}$y_{0}=0$\textbf{,
}$x_{0}=1$\textbf{).} \textbf{a,b)} Dependence of the meeting position
PDFs on $c$ for $f=0.4$. \textbf{c,d)} Dependence of the meeting
position PDFs on $f$ for $c=2.12$. The main figures show results
for $\nu=3/4$ while the insets those for $\nu=1/2$. The cusp from
the bubble centre PDFs (see Figure \ref{fig:06}) is projected into
the meeting position PDFs. The peak value increases with decreasing
$\nu$ (also compare with Figure \ref{fig:04}). The PDFs exhibit
universal power law asymptotics $y^{-3}$, which are independent of
$\nu$, $c$ and $f$ --- dotted lines.}
\label{fig:07} 
\end{figure}

These localisation properties can be used in the calculation and analysis
of the meeting position PDFs determined by the integral \eref{MeetingPositionDef}
\begin{equation}
\fl\pi_{\mathrm{mp}}^{(\nu)}(y|x_{0})=\int_{0}^{\infty}\mathrm{d}t\,\pi_{\mathrm{mt}}^{(\nu)}(t|x_{0})P_{\mathrm{c}}^{(\nu)}(y,t|0)=\int_{0}^{\infty}\frac{\mathrm{d}t}{t^{\frac{\nu}{2}}}\,\pi_{\mathrm{mt}}^{(\nu)}(t|x_{0})P_{c}^{(\nu)}\left(\frac{y}{t^{\frac{\nu}{2}}},1\bigg|0\right).
\end{equation}
Now, assuming $y>0$ (both $P_{c}^{(\nu)}(y,t|0)$ and, consequently,
also $\pi_{\mathrm{mp}}^{(\nu)}(y|x_{0})$ are even functions of $y$),
we substitute $t\to\eta=y/t^{\nu/2}$ in the above integral, which
yields
\begin{equation}
\pi_{\mathrm{mp}}^{(\nu)}(y|x_{0})=\frac{2}{\nu}y^{\frac{2}{\nu}-1}\int_{0}^{\infty}\mathrm{d}\eta\,\eta^{-\frac{2}{\nu}}P_{\mathrm{c}}^{(\nu)}(\eta,1|0)\pi_{\mathrm{mt}}^{(\nu)}\left(\left(\frac{y}{\eta}\right)^{\frac{2}{\nu}}\bigg|x_{0}\right).\label{eq:SubdiffMP}
\end{equation}
 This equation was used for the numerical evaluation of the meeting
position PDFs $\pi_{\mathrm{mp}}^{(\nu)}(y|x_{0})$ depicted in Figure~\ref{fig:07}.
Furthermore, the super-exponential localisation of the integrand factor
$\eta^{-2/\nu}P_{\mathrm{c}}^{(\nu)}(\eta,1|0)$ in Eq.~(\ref{eq:SubdiffMP})
around zero together with the power-law decay of $\pi_{\mathrm{mt}}^{(\nu)}\left(t\to\infty|x_{0}\right)\approx t^{-(1+\nu)}$
derived in Sec.~\ref{sub:Meeting-time-PDF} give a surprisingly universal
expression for the large-distance asymptotics 
\[
\pi_{\mathrm{mp}}^{(\nu)}(y\to\infty|x_{0})\approx y^{-3}\times\frac{2}{\nu}\intop_{0}^{\infty}d\eta\eta^{2}P_{c}^{(\nu)}(\eta,1|0),
\]
whose power exponent is just $-3$, completely independent of any
parameters of the model such as $f$, $c$, and, most remarkably,
also $\nu$. This universality is demonstrated in Figure~\ref{fig:07}
for the two cases with $\nu=1/2$ and $\nu=3/4$. 

\section{Dynamics in finite DNA chains\label{sec:Finite-chains}}

In a finite DNA molecule, the diffusive dynamics of bubble position
and its size depend on each other because the boundary conditions
break the translational invariance of the infinite-length problem
and, thus, connect the two variables. We need to consider a joint
PDF $P_{\mathrm{e}}(\xi,\chi,t|\xi_{0},\chi_{0})$ of the edge positions
$\xi$, $\chi$ that do not cross: $\xi<\chi$. We rescale $\xi$
and $\chi$ to take values between 0 and 1 (see the discussion on
scaling below). The bubble size $x$ and bubble position $y$ relate
to the edge locations through $\xi=y-x/2$ and $\chi=y+x/2$.

\subsection{Two-dimensional Fokker Planck equation}

Keeping the same form of the Gibbs free energy as in the infinite
chain case \eref{ContinuousGibbsFE}, the two-dimensional FPE reads
\begin{eqnarray}
\fl\left[-\partial_{t}+\partial_{\xi\xi}+\partial_{\chi\chi}+\left(\partial_{\chi}-\partial_{\xi}\right)\left(2f+\frac{2\gamma}{\chi-\xi}\right)\right]P_{\mathrm{e}}(\xi,\chi,t|\xi_{0},\chi_{0})=0\,,\quad\label{2DFPE}\\
P_{\mathrm{e}}(\xi,\chi,t\to0^{+}|\xi_{0},\chi_{0})=\delta(\xi-\xi_{0})\delta(\chi-\chi_{0})\,.\nonumber 
\end{eqnarray}
We track the closing dynamics of a single loop and impose the absorbing
boundary condition 
\begin{equation}
\lim_{\chi\rightarrow\xi^{+}}P_{\mathrm{e}}(\xi,\chi,t|\xi_{0},\chi_{0})=0\,.
\end{equation}
Bubble edges behave as \emph{vicious walkers} \cite{Fisher1984,BrayWinkler2004}
that annihilate each other upon meeting. This picture has already
helped investigating coalescence of DNA bubbles in Refs.~\cite{NovotnyPedersen2007,PedersenHansen2009}.

\begin{figure}[ht]
\centering \includegraphics{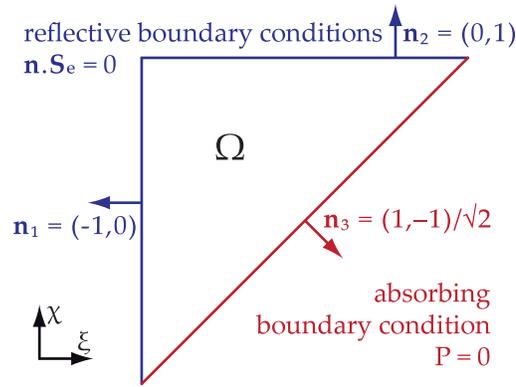} \protect\caption{\textbf{Support $\Omega$ and boundary conditions }of the finite-chain
bivariate Fokker-Planck partial differential equation (\ref{2DFPE}).
The reflective boundary conditions (blue legs) correspond to the ends
of the DNA chain where the strands are clamped together. The absorbing
boundary condition (red hypotenuse) corresponds to closing of the
bubble if $\xi=\chi$.}
\label{fig:08} 
\end{figure}

In our model, the single strands of DNA are considered to be clamped
together at the ends of the molecule. This assumption leads to the
reflecting boundary conditions 
\begin{eqnarray}
\left[\partial_{\xi}-2f-\frac{2\gamma}{\chi-\xi}\right]P_{\mathrm{e}}(\xi,\chi,t|\xi_{0},\chi_{0})\Big|_{\xi=0} & =0\,,\\
\left[\partial_{\chi}+2f+\frac{2\gamma}{\chi-\xi}\right]P_{\mathrm{e}}(\xi,\chi,t|\xi_{0},\chi_{0})\Big|_{\chi=1} & =0\,.
\end{eqnarray}
Admissible positions of the edges form an isosceles right-angled triangle
with the absorbing condition on the hypotenuse and the reflecting
conditions on the legs, described as $\Omega$ in Figure \ref{fig:08}.

The effect of the entropic term $\gamma/x$ relative to the energetic
term $f$ is strongest for small bubbles or for low $f$, i.e., close
to the melting temperature. The equations \eref{FPEgeneral}, \eref{FFPEbubbleSize},
and \eref{2DFPE} can be rescaled as follows 
\begin{eqnarray}
x & =Ax'\nonumber \\
f & =f'/A\nonumber \\
\gamma & =\gamma'\label{Scaling}\\
t & =A^{2/\nu}t'\,.\nonumber 
\end{eqnarray}
Note that the loop exponent $c=2\gamma$ stays unchanged under the
scaling. Position variables $y$, $\xi$, and $\chi$ transform in
the same fashion as $x.$ The principal relation is $fx=f'x'$, because
the scaling of the time variable can be absorbed into the fitting
of the diffusion constant. The measure $P_{\mathrm{e}}(\xi,\chi,t|\xi_{0},\chi_{0})\mathrm{d}\xi\mathrm{d}\chi$ 
 stays invariant in order to ensure a consistent norming of probability,
yielding $P_{\mathrm{e}}(\xi,\chi,t|\xi_{0},\chi_{0})=A^{-2}P_{\mathrm{e}}(\xi',\chi',t'|\xi'_{0},\chi'_{0})$ 
, from which the scaling of the initial and boundary conditions follows
(compare also with the discussion below Eq.~(\ref{SubdiffFPEcentre})).

\subsection{Time-integrated quantities}

We solve the dynamics in finite chains numerically. Derived quantities
-- the first moment of the meeting time and the meeting position PDF
-- need only time-integrated information, which reduces the numerical
effort.

\subsubsection{Moments of the meeting time PDF.}

Closing of a bubble reduces the proportion of bubbles surviving; the
meeting time PDF equals the negative of the time derivative of the
spatial integral of $P_{\mathrm{e}}(\xi,\chi,t|\xi_{0},\chi_{0})$
over $\Omega$ \cite[Sec.~8.1]{Risken1989} 
\begin{eqnarray}
\pi_{\mathrm{mt}}(t|\xi_{0},\chi_{0}) & =-\frac{\mathrm{d}}{\mathrm{d}t}\int_{\Omega}\mathrm{d}\xi\mathrm{d}\chi P_{\mathrm{e}}(\xi,\chi,t|\xi_{0},\chi_{0})=\\
 & =-\int_{0}^{1}\mathrm{d}\xi\int_{\xi}^{1}\mathrm{d}\chi\frac{\mathrm{d}P_{\mathrm{e}}(\xi,\chi,t|\xi_{0},\chi_{0})}{\mathrm{d}t}\,.\nonumber 
\end{eqnarray}
The first moment of the meeting time PDF is defined by 
\begin{eqnarray}
\fl T_{1}(\xi_{0},\chi_{0}) & =\int_{0}^{\infty}t\pi_{\mathrm{mt}}(t|\xi_{0},\chi_{0})\mathrm{d}t=\int_{\Omega}p_{1}(\xi,\chi|\xi_{0},\chi_{0})\mathrm{d}\Sigma\quad,\,\mathrm{where}\\
\fl p_{1}(\xi,\chi|\xi_{0},\chi_{0}) & =-\int_{0}^{\infty}t\frac{\mathrm{d}P_{\mathrm{e}}(\xi,\chi,t|\xi_{0},\chi_{0})}{\mathrm{d}t}\mathrm{d}t=\int_{0}^{\infty}P_{\mathrm{e}}(\xi,\chi,t|\xi_{0},\chi_{0})\mathrm{d}t\,.
\end{eqnarray}
We apply $\mathbb{L}_{\mathrm{FP}}$ -- the spatial part of \eref{2DFPE}
-- on both sides of the previous equation and use that $P_{\mathrm{e}}(\xi,\chi,t|\xi_{0},\chi_{0})$
solves the FPE $\mathbb{L}_{\mathrm{FP}}P_{\mathrm{e}}(\xi,\chi,t|\xi_{0},\chi_{0})=\partial_{t}P_{\mathrm{e}}(\xi,\chi,t|\xi_{0},\chi_{0})$
with the $\delta$-function initial condition (\ref{2DFPE}) \cite[Sec.~8.1]{Risken1989}
\begin{equation}
\mathbb{L}_{\mathrm{FP}}p_{1}(\xi,\chi|\xi_{0},\chi_{0})=-\delta(\xi-\xi_{0})\delta(\chi-\chi_{0})\,.
\end{equation}
We have determined the auxiliary function $p_{1}(\xi,\chi|\xi_{0},\chi_{0})$
numerically using the finite element method (FEM), which has the advantage
that $\delta$-distribution has a simple implementation in the weak
formulation of the FPE. A hierarchy of auxiliary functions exists
for higher moments of $\pi_{\mathrm{mt}}(t|\xi_{0},\chi_{0})$ \cite[Sec.~8.1]{Risken1989}
and the approach using FEM can be suitably generalised.

\begin{figure}[ht]
\centering \includegraphics{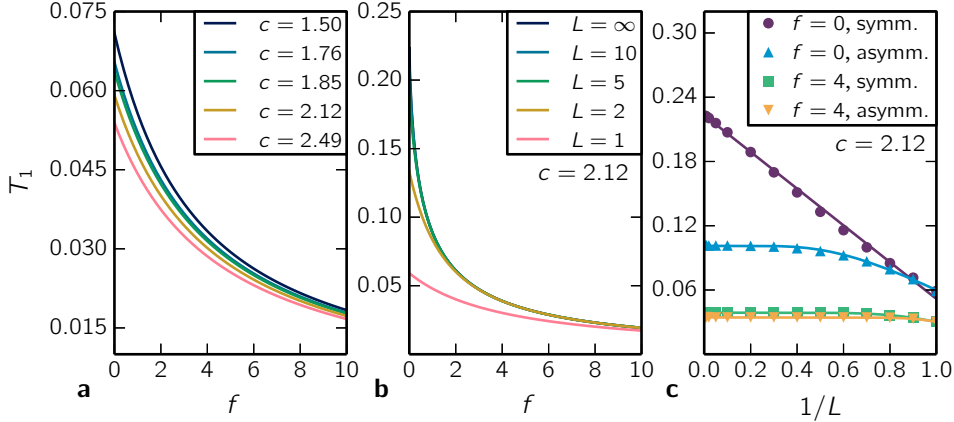} \protect\caption{\textbf{Mean meeting time --- finite chain.} \textbf{a)} Dependence
of the mean meeting time $T_{1}$ on $c$ and $f$ for $L=1$, $\xi_{0}=0$,
and $\chi_{0}=1$. \textbf{b)} Scaling of $T_{1}$ for a bubble of
initial size $x_{0}=1$ in a chain of length $L$ in dependence on
$f$. The infinite chain results are approached exponentially in case
of finite $f$. \textbf{c)} Comparison of a bubble placed symmetrically
in the centre of the chain with a bubble placed asymmetrically at
its edge. Presence of the boundary causes the exponential scaling
to remain at the melting temperature ($f=0$), while the centrally
placed bubble exhibits scaling linear with $L^{-1}$. Full lines are
fits of linear $T_{1}(\infty)+(T_{1}(1)-T_{1}(\infty))/L$ or exponential
$T_{1}(\infty)+(T_{1}(1)-T_{1}(\infty))\exp(-(L-1)/\ell)$ functions,
respectively, with $T_{1}(\infty)$ being analytically known in the
case of the symmetric initial condition or just another fitting parameter
in the asymmetric case. The obtained values for the characteristic
convergence length are $\ell(\blacktriangle)=0.42\pm0.02$, $\ell(\blacksquare)=0.192\pm0.001$,
and $\ell(\blacktriangledown)=0.096\pm0.006$.}
\label{fig:09} 
\end{figure}

\subsubsection{Mean meeting time}

In Figure \ref{fig:09}, we compare the numerical results for $T_{1}$
in the case of finite chains with the infinite chain mean meeting
time $T_{1}(\infty)$ obtained from Eqs.~(\ref{MeanMeetingTimeDef})
and (\ref{DiffusiveMeetingTime}). We rescaled the results to keep
the initial size of the bubble fixed at $x_{0}=1$, whereas we vary
the total chain length $L$, using the scaling relations (\ref{Scaling}).
We observe a rapid exponential convergence with a characteristic convergence
length $\ell\ll1$ to the infinite chain value, leading to $|T_{1}(10)/T_{1}(\infty)-1|<\epsilon_{\mathrm{FEM}}\simeq10^{-5}$,
thus being indistinguishable for the FEM implementation used here.
The sole exception is the critical point, where $T_{1}(\infty)-T_{1}(L)\propto L^{-1}$.
This is consistent with the entropic term being scale-free as shown
in Eq.~(\ref{Scaling}). The exponential scaling can be recovered
by initially placing the bubble in the vicinity of one of the chain
ends. Generally, the presence of ends reduces the mean meeting time
as it prevents the bubble from spreading out as is demonstrated in
Figure \ref{fig:10}. 

\subsubsection{Meeting position PDF}

\begin{figure}[ht]
\centering \includegraphics{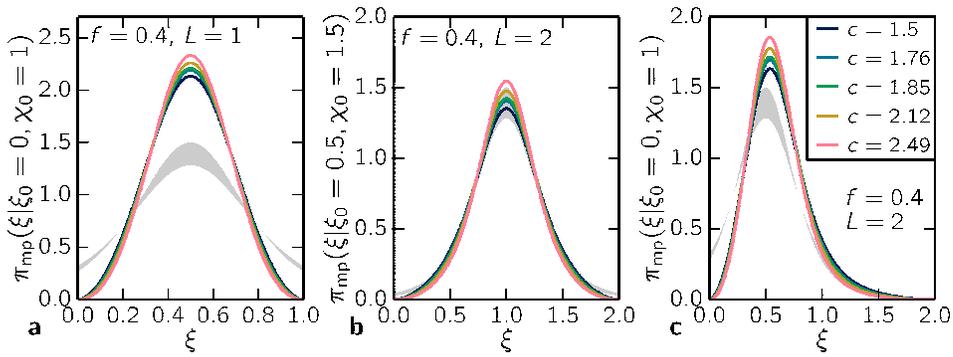} \protect\caption{\textbf{Meeting position --- finite chain.} Meeting position PDF for
several values of $c$ for $f=0.4$ for different initial conditions:
\textbf{a)} $L=1$, $\xi_{0}=0$ and $\chi_{0}=1$, \textbf{b)} $L=2$,
$\xi_{0}=0.5$ and $\chi_{0}=1.5$. The PDF rapidly approaches the
infinite chain value in case of the symmetric initial placement of
the bubble. \textbf{c)} $L=2$, $\xi_{0}=0$ and $\chi_{0}=1$. The
presence of the boundary modifies the shape of the distribution, which
becomes more concentrated around the peak values. The shaded area
represents the range of infinite chain values for identical parameters.}
\label{fig:10} 
\end{figure}

The meeting position PDF is the time integrated probability that the
bubble closes at a given position $\xi=\chi$ along the absorbing
boundary. We restate it as the time-integrated projection of the probability
current $\vec{S}_{\mathrm{e}}$ to the direction $\vec{n}_{3}$ orthogonally
crossing the absorbing boundary in Fig.~\ref{fig:08} taking on the
following form 
\begin{eqnarray}
\fl\left[\vec{S}_{\mathrm{e}}(\xi,\chi,t|\xi_{0},\chi_{0})\cdot\vec{n_{3}}\right]_{\chi\rightarrow\xi^{+}} & =\Bigg[\left(-\partial_{\xi}+2f+\frac{2\gamma}{\chi-\xi},-\partial_{\chi}-2f-\frac{2\gamma}{\chi-\xi}\right)P_{\mathrm{e}}(\xi,\chi,t|\xi_{0},\chi_{0})\cdot\\
 & \cdot\frac{1}{\sqrt{2}}(1,-1)\Bigg]_{\chi\rightarrow\xi^{+}}=\frac{\sqrt{2}}{2}\left[\left(-\partial_{\xi}+\partial_{\chi}+\frac{4\gamma}{\chi-\xi}\right)P_{\mathrm{e}}(\xi,\chi,t|\xi_{0},\chi_{0})\right]_{\chi\rightarrow\xi^{+}}\,,\nonumber 
\end{eqnarray}
where we have used the absorbing boundary condition to eliminate the
term containing $f$. The time integral of the previous equation is
\begin{eqnarray}
\fl\pi_{\mathrm{mp}}(\xi|\xi_{0},\chi_{0}) & =\int_{0}^{\infty}\mathrm{d}t\left[\vec{S}_{e}(\xi,\chi,t|\xi_{0},\chi_{0})\cdot\vec{n_{3}}\right]_{\chi\rightarrow\xi^{+}}=\nonumber \\
 & =\frac{1}{\sqrt{2}}\int_{0}^{\infty}\mathrm{d}t\left[\left(-\partial_{\xi}+\partial_{\chi}+\frac{4\gamma}{\chi-\xi}\right)P_{e}(\xi,\chi,t|\xi_{0},\chi_{0})\right]_{\chi\rightarrow\xi^{+}}=\nonumber \\
 & =\frac{1}{\sqrt{2}}\left[\left(-\partial_{\xi}+\partial_{\chi}+\frac{4\gamma}{\chi-\xi}\right)p_{1}(\xi,\chi|\xi_{0},\chi_{0})\right]_{\chi\rightarrow\xi^{+}}\,.
\end{eqnarray}
Therefore the meeting position PDF can also be obtained from the first
auxiliary function $p_{1}(\xi,\chi|\xi_{0},\chi_{0})$ as the first
moment of the meeting time PDF above. Numerical observations show
that both partial derivatives and $p_{1}(\xi,\chi|\xi_{0},\chi_{0})/(\chi-\xi)$
stay finite and non-zero as the limit $\chi\rightarrow\xi^{+}$ is
approached.\textbf{ }Figure \ref{fig:10} shows results for selected
parameter values. As in the case of the mean meeting time, we observe
a rapid convergence of the meeting position PDF to the infinite chain
values if the bubble is placed symmetrically. Placing the bubble at
the extremity of the chain significantly modifies the shape of the
PDF leading to a sharper peak around the initial position of the bubble
centre. 

\section{Possible experimental verifications\label{sec:Experimental-verification}}

Successful experimental observation requires careful choice of the
initial bubble size and the measurement temperature. Increasing temperature
introduces more experimental noise into the measurements. In heteropolymers,
the subdiffusive regime appears at temperatures close to the melting
temperature. On the other hand, at low temperatures, the entropic
effects play a reduced role.

The force $f=\beta\Delta G_{\mathrm{bp}}/2$ resulting from the Gibbs
free energy can be expressed in terms of the enthalpy and entropy
differences 
\begin{equation}
f=\frac{\Delta H(T)-T\Delta S(T)}{2k_{B}T}\,.
\end{equation}
At critical temperature $\Delta H(T_{c})=T_{c}\Delta S(T_{c})$; we
linearly approximate the drift term 
\begin{equation}
f=\frac{\Delta S(T_{c})(T_{c}-T)}{2k_{B}T}=\frac{\Delta S^{\circ}(T_{c})(T_{c}-T)}{2RT}\,.
\end{equation}
($R=k_{B}N_{A}\doteq8.31\mathrm{J/(K.mol)}$ is the gas constant and
$N_{A}$ the Avogadro number). An example DNA chain with average molar
$\Delta H^{\circ}(T_{c})\equiv N_{A}\Delta H(T_{c})=30\mathrm{kJ/mol}$
and $\Delta S^{\circ}(T_{c})\equiv N_{A}\Delta S(T_{c})=90\mathrm{J/(K.mol)}$
has $T_{c}=333\mathrm{K}$. The force at physiological temperature
is thus $f\simeq0.4$. 

Previous experimental studies have measured the auto-correlation function
which can be related to the bubble survival probability. The auto-correlation
function is an integral of the meeting time PDF. In the case of the
meeting time PDFs, we treated the diffusion constant as a parameter
which would first have to be determined independently or fitted. Instead,
we focus on the meeting position PDF in discussing the possible experimental
realisations. The meeting position PDF is independent of the exact
value of the diffusion constant, as the time variable is integrated
out.

\begin{figure}[ht]
\centering \includegraphics{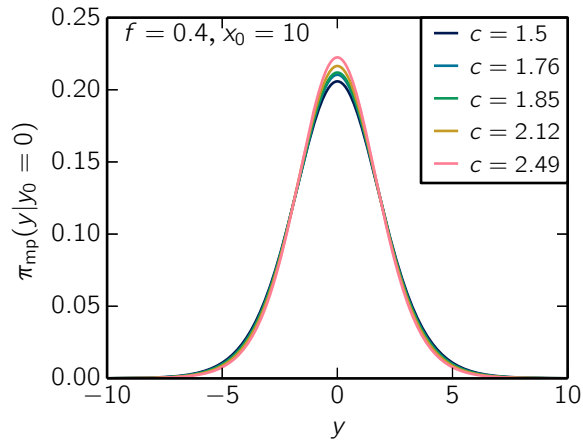} \protect\caption{\textbf{Possibility of experimental observation.} For a DNA homopolymer
around physiological temperatures and initial bubble size of 10 base
pairs, the relative difference between the peak values of the meeting
position PDFs for $c=2.12$ and $c=1.76$ is 2.8~\%.}
\label{fig:11} 
\end{figure}

The difference between various values of $c$ can be clearly discerned
at the peak as opposed to the tails, underlining the importance of
knowing the entire meeting position PDF. The difference in the peak
value is 2.8 \% between $c=2.12$ and $c=1.76$ for a bubble of initial
size of ten base pairs (see Figure \ref{fig:11}). For a bubble of
initial size of 100 bp and 1000 bp, the difference reduces to 0.68
\% and 0.17 \%, respectively. Our method can be extended to cases
when the initial bubble size has a known distribution; this requires
a convolution of the distribution with the Green function. An experimental
observation would require a preparation of bubbles of sizes of tens
of base pairs at physiological temperatures and subsequent observation
of the histogram of their closing position with a precision of $10^{-2}$.

Moreover, our results show a rapid approach of the finite chain to
the infinite chain results. Therefore, precise knowledge of the chain
length should not be crucial and a chain only several times longer
than the initial bubble size would be sufficient to obtain data described
by the infinite chain PDFs.

The approach in this work reduces the three-dimensional conformational
dynamics of the DNA chain to the effective one-dimensional model with
entropic interactions. Such dimensional reduction appears leads to
logarithmic terms in free energy in other models as well. The interplay
with bending and twisting of the DNA chain in three dimensions may
further influence the bubble dynamics --- e.g., by formation of a
bend in the bubble which prevents its closing \cite{Strick:PNAS98}.
Care would have to be taken to separate the closing dynamics as described
here from such additional events. For example, a light stretching
force would reduce the formation of bends.

\section{Conclusions\label{sec:Conclusions}}

We have presented a solution of the Fokker-Planck equation governing
DNA bubble dynamics in the framework of the Poland-Scheraga model
both for normal diffusion, as well as for its fractional counterpart
describing the subdiffusive dynamics in heteropolymers.

Expressions for the meeting time and meeting position PDFs have been
given in the analytical form in the Laplace/Fourier image and numerically
transformed back to the full time/position variable range, thus extending
previous results focused mainly on the asymptotic behaviour of the
meeting time PDF. Meeting position PDF, which is a promising newly
suggested quantity for the determination of the value of the entropic
loop exponent, shows in the subdiffusive case surprisingly universal
asymptotic features which appear to be pertinent to a whole class
of stochastic processes. Furthermore, we have studied the influence
of a finite length of the DNA chain on the diffusive dynamic of a
bubble and shown a generic (with the only exception for the symmetrically
placed bubble at the critical point) exponentially fast convergence
to the infinite length limit. 

Our solution constitutes a reference point for more involved models
or numerical simulations. The results are relevant for determining
the value of the entropic loop exponent from the bubble breathing
dynamics in future experiments.

\ack{}{We thank Prof.~Ralf Metzler for useful discussions and hospitality
during our visits to his group. V.K.~thanks Jens Karschau and Malte
Vogl for valuable remarks on the manuscript. This work was financially
supported by the Czech Science Foundation through grant No.~P205/10/0989
(T.N.).}

\section*{References}

\bibliographystyle{unsrt}
\bibliography{bubbles}

\begin{thebibliography}{10}

\bibitem{Ambjornsson:BiophysJ07}
Tobias Ambj{\"o}rnsson, Suman~K. Banik, Oleg Krichevsky, and Ralf Metzler.
\newblock Breathing dynamics in heteropolymer {DNA}.
\newblock {\em Biophysical Journal}, 92(8):2674--2684, Apr 2007.

\bibitem{Peyrard:NatPhys06}
Michel Peyrard.
\newblock Biophysics: Melting the double helix.
\newblock {\em Nature Physics}, 2(1):13--14, Jan 2006.

\bibitem{PolandScheraga1966A}
Douglas Poland and Harold~A. Scheraga.
\newblock Phase transitions in one dimension and the helix--coil transition in
  polyamino acids.
\newblock {\em The Journal of Chemical Physics}, 45(5):1456--1463, 1966.

\bibitem{PolandScheraga1966B}
Douglas Poland and Harold~A. Scheraga.
\newblock Occurrence of a phase transition in nucleic acid models.
\newblock {\em The Journal of Chemical Physics}, 45(5):1464--1469, 1966.

\bibitem{Barkai:PRE12}
A.~Dechant, E.~Lutz, D.~A. Kessler, and E.~Barkai.
\newblock Superaging correlation function and ergodicity breaking for
  {B}rownian motion in logarithmic potentials.
\newblock {\em Physical Review E}, 85(5):051124, May 2012.

\bibitem{Mukamel:JSTAT12}
Ori Hirschberg, David Mukamel, and Gunter~M Sch{\"u}tz.
\newblock Diffusion in a logarithmic potential: scaling and selection in the
  approach to equilibrium.
\newblock {\em Journal of Statistical Mechanics: Theory and Experiment},
  2012(02):P02001, 2012.

\bibitem{HankeMetzler2003}
Andreas Hanke and Ralf Metzler.
\newblock Bubble dynamics in {DNA}.
\newblock {\em Journal of Physics A: Mathematical and General}, 36(36):L473,
  2003.

\bibitem{Altan-Bonnet2003}
Gr\'egoire Altan-Bonnet, Albert Libchaber, and Oleg Krichevsky.
\newblock Bubble dynamics in double-stranded {DNA}.
\newblock {\em Phys. Rev. Lett.}, 90:138101, Apr 2003.

\bibitem{FogedbyMetzler2007A}
Hans~C. Fogedby and Ralf Metzler.
\newblock {DNA} bubble dynamics as a quantum {C}oulomb problem.
\newblock {\em Phys. Rev. Lett.}, 98:070601, Feb 2007.

\bibitem{FogedbyMetzler2007B}
Hans~C. Fogedby and Ralf Metzler.
\newblock Dynamics of {DNA} breathing: Weak noise analysis, finite time
  singularity, and mapping onto the quantum {C}oulomb problem.
\newblock {\em Phys. Rev. E}, 76:061915, Dec 2007.

\bibitem{Bar:JPCM09}
A~Bar, Y~Kafri, and D~Mukamel.
\newblock Dynamics of {DNA} melting.
\newblock {\em Journal of Physics: Condensed Matter}, 21(3):034110, 2009.

\bibitem{Ambjornsson:PRL06}
Tobias Ambj{\"o}rnsson, Suman~K. Banik, Oleg Krichevsky, and Ralf Metzler.
\newblock Sequence sensitivity of breathing dynamics in heteropolymer {DNA}.
\newblock {\em Physical Review Letters}, 97(12):128105, Sep 2006.

\bibitem{Jeon:JCP06}
Jae-Hyung Jeon, Pyeong~Jun Park, and Wokyung Sung.
\newblock The effect of sequence correlation on bubble statistics in
  double-stranded {DNA}.
\newblock {\em The Journal of Chemical Physics}, 125(16):164901, 2006.

\bibitem{Banik:JCP11}
Srijeeta Talukder, Pinaki Chaudhury, Ralf Metzler, and Suman~K. Banik.
\newblock Determining the {DNA} stability parameters for the breathing dynamics
  of heterogeneous {DNA} by stochastic optimization.
\newblock {\em The Journal of Chemical Physics}, 135(16):165103, 2011.

\bibitem{BenhamBi2004}
Craig~J. Benham and Chengpeng Bi.
\newblock The analysis of stress-induced duplex destabilization in long genomic
  {DNA} sequences.
\newblock {\em Journal of Computational Biology}, 11:519--543, 2004.

\bibitem{JeonAdamcik2010}
Jae-Hyung Jeon, Jozef Adamcik, Giovanni Dietler, and Ralf Metzler.
\newblock Supercoiling induces denaturation bubbles in circular {DNA}.
\newblock {\em Phys. Rev. Lett.}, 105:208101, Nov 2010.

\bibitem{Ma:Science13}
Jie Ma, Lu~Bai, and Michelle~D. Wang.
\newblock Transcription under torsion.
\newblock {\em Science}, 340(6140):1580--1583, Jun 2013.

\bibitem{King:PNAS13}
Graeme~A. King, Peter Gross, Ulrich Bockelmann, Mauro Modesti, Gijs J.~L.
  Wuite, and Erwin J.~G. Peterman.
\newblock Revealing the competition between peeled ss{DNA}, melting bubbles,
  and {S-DNA} during {DNA} overstretching using fluorescence microscopy.
\newblock {\em Proceedings of the National Academy of Sciences},
  110(10):3859--3864, Mar 2013.

\bibitem{Oberstrass:PRL13}
Florian~C. Oberstrass, Louis~E. Fernandes, Paul Lebel, and Zev Bryant.
\newblock Torque spectroscopy of {DNA}: Base-pair stability, boundary effects,
  backbending, and breathing dynamics.
\newblock {\em Physical Review Letters}, 110(17):178103, Apr 2013.

\bibitem{Fisher1966}
Michael~E. Fisher.
\newblock Effect of excluded volume on phase transitions in biopolymers.
\newblock {\em The Journal of Chemical Physics}, 45(5):1469--1473, 1966.

\bibitem{KafriMukamel2000}
Yariv Kafri, David Mukamel, and Luca Peliti.
\newblock Why is the {DNA} denaturation transition first order.
\newblock {\em Phys. Rev. Lett.}, 85:4988--4991, Dec 2000.

\bibitem{BlosseyCarlon2003}
Ralf Blossey and Enrico Carlon.
\newblock Reparametrizing the loop entropy weights: Effect on {DNA} melting
  curves.
\newblock {\em Phys. Rev. E}, 68:061911, Dec 2003.

\bibitem{Wu:PRE09}
Lian-Ao Wu, Stephen~S. Wu, and Dvira Segal.
\newblock Looking into {DNA} breathing dynamics via quantum physics.
\newblock {\em Physical Review E}, 79(6):061901, Jun 2009.

\bibitem{HwaMarinari2003}
Terence Hwa, Enzo Marinari, Kim Sneppen, and Lei-han Tang.
\newblock Localization of denaturation bubbles in random {DNA} sequences.
\newblock {\em Proceedings of the National Academy of Sciences},
  100(8):4411--4416, 2003.

\bibitem{Bicout:PRE04}
D.~J. Bicout and E.~Kats.
\newblock Bubble relaxation dynamics in double-stranded {DNA}.
\newblock {\em Physical Review E}, 70(1):010902, Jul 2004.

\bibitem{NovotnyPedersen2007}
Tom\'{a}\v{s} Novotn\'{y}, Jonas~Nyvold Pedersen, Tobias Ambj\"{o}rnsson,
  Mikael~Sonne Hansen, and Ralf Metzler.
\newblock Bubble coalescence in breathing {DNA}: Two vicious walkers in
  opposite potentials.
\newblock {\em EPL (Europhysics Letters)}, 77(4):48001, 2007.

\bibitem{PedersenHansen2009}
Jonas~Nyvold Pedersen, Mikael~Sonne Hansen, Tom\'{a}\v{s} Novotn\'{y}, Tobias
  Ambj\"{o}rnsson, and Ralf Metzler.
\newblock Bubble merging in breathing {DNA} as a vicious walker problem in
  opposite potentials.
\newblock {\em The Journal of Chemical Physics}, 130(16):164117, 2009.

\bibitem{KafriMukamel2002}
Y.~Kafri, D.~Mukamel, and L.~Peliti.
\newblock Melting and unzipping of {DNA}.
\newblock {\em The European Physical Journal B - Condensed Matter and Complex
  Systems}, 27:135--146, 2002.

\bibitem{SantaLuciaHicks2004}
John SantaLucia and Donald Hicks.
\newblock The thermodynamics of {DNA} structural motifs.
\newblock {\em Annual Review of Biophysics and Biomolecular Structure},
  33(1):415--440, 2004.

\bibitem{HouchesLRI}
D.~Mukamel.
\newblock Statistical mechanics of systems with long-range interactions.
\newblock In Thierry Dauxois, Stefano Ruffo, and Leticia~F. Cugliandolo,
  editors, {\em Long-Range Interacting Systems: Lecture Notes of the Les
  Houches Summer School}, volume~90. Oxford University Press, 2010.

\bibitem{Risken1989}
Hannes Risken.
\newblock {\em {F}okker-{P}lanck equation: {M}ethods of Solution and
  Applications}.
\newblock Springer-Verlag, Berlin, 2nd edition, 1989.

\bibitem{NISTfunctions2010}
Frank W.~J. Olver, Daniel~W. Lozier, Ronald~F. Boisvert, and Charles~W. Clark.
\newblock {\em NIST Handbook of Mathematical Functions}.
\newblock Cambridge University Press, Cambridge, 2010.

\bibitem{Redner2001}
Sidney Redner.
\newblock {\em A Guide to First-Passage Processes}.
\newblock Cambridge University Press, 2001.

\bibitem{Novotny:PRE00}
Tom{\'a}{\v s} Novotn{\'y} and Petr Chvosta.
\newblock Resonant activation phenomenon for non-{M}arkovian
  potential-fluctuation processes.
\newblock {\em Physical Review E}, 63(1):012102, Dec 2000.

\bibitem{AbateWhitt2006}
Joseph Abate and Ward Whitt.
\newblock A unified framework for numerically inverting {L}aplace transforms.
\newblock {\em INFORMS J. on Computing}, 18(4):408--421, January 2006.

\bibitem{HupperPollak1999}
Bruno Hupper and Eli Pollak.
\newblock Numerical inversion of the {L}aplace transform.
\newblock {\em The Journal of Chemical Physics}, 110(23):11176--11186, 1999.

\bibitem{Stehfest1970}
Harald Stehfest.
\newblock Algorithm 368: {N}umerical inversion of {L}aplace transforms {[D5]}.
\newblock {\em Commun. ACM}, 13(1):47--49, January 1970.

\bibitem{Hanke:PRL08}
Andreas Hanke, Martha~G. Ochoa, and Ralf Metzler.
\newblock Denaturation transition of stretched {DNA}.
\newblock {\em Physical Review Letters}, 100(1):018106, Dec 2008.

\bibitem{Marenduzzo:PRE10}
D.~Marenduzzo, E.~Orlandini, F.~Seno, and A.~Trovato.
\newblock Different pulling modes in {DNA} overstretching: A theoretical
  analysis.
\newblock {\em Physical Review E}, 81(5):051926, May 2010.

\bibitem{Netz:PRL08}
Thomas~R. Einert, Paul N{\"a}ger, Henri Orland, and Roland~R. Netz.
\newblock Impact of loop statistics on the thermodynamics of {RNA} folding.
\newblock {\em Physical Review Letters}, 101(4):048103, Jul 2008.

\bibitem{PressTeukolsky2007}
William~H. Press, Saul~A. Teukolsky, William~T. Vetterling, and Brian~P.
  Flannery.
\newblock {\em Numerical Recipes: The Art of Scientific Computing}.
\newblock Cambridge University Press, Cambridge, 3rd edition, 2007.

\bibitem{MetzlerKlafter2000}
Ralf Metzler and Joseph Klafter.
\newblock The random walk's guide to anomalous diffusion: a fractional dynamics
  approach.
\newblock {\em Physics Reports}, 339(1):1 -- 77, 2000.

\bibitem{MetzlerKlafter2004}
Ralf Metzler and Joseph Klafter.
\newblock The restaurant at the end of the random walk: recent developments in
  the description of anomalous transport by fractional dynamics.
\newblock {\em Journal of Physics A: Mathematical and General}, 37(31):R161,
  2004.

\bibitem{Feller1971}
William Feller.
\newblock {\em An Introduction to Probability Theory and Its Applications}.
\newblock John Wiley \& Sons, 1971.

\bibitem{Lua:PRE05}
Rhonald~C. Lua and Alexander~Y. Grosberg.
\newblock First passage times and asymmetry of {DNA} translocation.
\newblock {\em Physical Review E}, 72(6):061918, Dec 2005.

\bibitem{Condamin:PRL07}
S.~Condamin, O.~B{\'e}nichou, and J.~Klafter.
\newblock First-passage time distributions for subdiffusion in confined
  geometry.
\newblock {\em Physical Review Letters}, 98(25):250602, Jun 2007.

\bibitem{Palyulin:JPA14}
Vladimir~V Palyulin and Ralf Metzler.
\newblock Speeding up the first-passage for subdiffusion by introducing a
  finite potential barrier.
\newblock {\em Journal of Physics A: Mathematical and Theoretical},
  47(3):032002, 2014.

\bibitem{Fisher1984}
Michael~E. Fisher.
\newblock Walks, walls, wetting, and melting.
\newblock {\em Journal of Statistical Physics}, 34:667--729, 1984.
\newblock 10.1007/BF01009436.

\bibitem{BrayWinkler2004}
Alan~J Bray and Karen Winkler.
\newblock Vicious walkers in a potential.
\newblock {\em Journal of Physics A: Mathematical and General}, 37(21):5493,
  2004.

\bibitem{Strick:PNAS98}
T.~R. Strick, V.~Croquette, and D.~Bensimon.
\newblock Homologous pairing in stretched supercoiled {DNA}.
\newblock {\em Proceedings of the National Academy of Sciences},
  95(18):10579--10583, Sep 1998.

\end{thebibliography}

\end{document}